\documentstyle[11pt,paspconf,epsf]{article}

\begin{document}

\title{The Second-Parameter Effect in Metal-Rich Globular Clusters}
\author{Allen V. Sweigart}
\affil{NASA Goddard Space Flight Center, Code 681, Greenbelt, MD 20771}

\begin{abstract}
Recent Hubble Space Telescope (HST) observations by Rich et al. (1997)
have shown that the metal-rich globular clusters (GCs) NGC 6388 and NGC 6441
exhibit a pronounced 2nd parameter effect.  Ordinarily metal-rich GCs have
only a red horizontal-branch (HB) clump.  However, NGC 6388 and NGC 6441
also possess an unexpected population of blue HB stars, indicating that
some 2nd parameter is operating in these
clusters.  Quite remarkably, the HBs in both clusters slope upward
with decreasing {\bv} from the red clump to the top of the blue tail.
     
We review the results of ongoing stellar evolution calculations
which indicate 1) that NGC 6388 and NGC 6441 might provide a crucial
diagnostic for understanding the origin of the
2nd parameter effect, 2) that differences in age or mass loss along
the red-giant branch (RGB) - the two most prominent 2nd parameter
candidates - cannot explain the HB morphology of these GCs, and 3) that
noncanonical effects involving an enhanced helium abundance or
rotation can produce upward sloping HBs.  Finally
we suggest a new metal-depletion scenario which might help
to resolve a baffling conundrum concerning the surface gravities
of the blue HB stars in these clusters.
\end{abstract}


\section{Introduction}

Ever since its discovery over 30 yrs ago by Sandage \& Wildey (1967) and
van den Bergh (1967), the 2nd parameter effect has stood as one of the
major unsolved challenges in the study of the Galactic GCs.  Although
it was recognized quite early that the HB morphology becomes redder
on average with increasing metallicity (Sandage \& Wallerstein 1960),
many pairs of GCs were also known with identical metallicities
but markedly different HB morphologies, e.g., M 3 versus M 13, NGC 288
versus NGC 362.  Indeed, some GCs, e.g., NGC 1851 and NGC 2808, possess
bimodal HBs, where the 2nd parameter effect seems to be producing
both blue and red HB populations within the same cluster.  Thus
some parameter(s) besides metallicity (the 1st parameter) must
be affecting the evolution of the HB stars in these GCs.
  
Many 2nd parameter candidates
have been suggested over the years, including
the GC age, mass loss along the RGB, helium abundance $Y$, alpha-element
abundance [$\alpha$/Fe], cluster dynamics, stellar rotation, deep mixing,
etc. (see Fusi Pecci \& Bellazzini 1997 for a recent review).  Determining
which of these candidates is responsible for the 2nd parameter effect
has been a difficult, but exceedingly important, task for
understanding a number of astrophysical problems.  In
particular, the age interpretation of the 2nd
parameter effect requires age differences of several Gyr among
the GCs, implying a correspondingly long timescale for the formation
of the Galactic halo.  Recent age determinations from both the
``horizontal'' and ``vertical'' methods do not, however, support such
large age differences, indicating that age is probably not the dominant
2nd parameter except perhaps in a few specific cases (see, e.g., VandenBerg
1999).  The 2nd parameter effect is also important for understanding
the origin of the extreme HB stars which are found in several
GCs (e.g., M 13, NGC 2808, NGC 6752, $\omega$ Cen) and which are
believed to be responsible for the UV upturn phenomenon in
elliptical galaxies.  Why some GCs possess such hot HB stars
while others do not remains an open puzzle.  Finally
the 2nd parameter effect
provides a potential means for assessing the impact of such noncanonical
processes as cluster dynamics, rotation and mixing on the evolution
of GC stars.
             
The HST observations of NGC 6388 and NGC 6441 by Rich et al. (1997)
first demonstrated that old, metal-rich stellar populations can
produce hot HB stars.  As noted by Sweigart \& Catelan (1998,
hereafter SC98), the HBs in these clusters
slope upward with decreasing {\bv} with the luminosity at the top of the
blue tail being nearly 0.5 mag brighter in $V$ than the well-populated
red HB clump.  This pronounced upward slope to the HB provides a crucial
diagnostic for the 2nd parameter effect because it immediately tells us
that the 2nd parameter is affecting both the temperature and luminosity
of the HB stars.  Smaller upward slopes of ${\approx} 0.1$
mag have also been reported in the bimodal clusters NGC 1851 (Walker
1998) and NGC 6229 (Borissova et al. 1999), which interestingly
are among the more metal-rich of the intermediate-metallicity
GCs.
     
The fact that hot HB stars are found in GCs as metal-rich as NGC 6388
and NGC 6441 (${\rm [Fe/H]} \, \approx \, -0.5$) may be especially important
for the following reason.  In intermediate-metallicity GCs such as M 3
the HB spans a wide range in color that extends both blueward and
redward of the instability strip.  The location of a star along the HB
is then quite sensitive to changes in the stellar
parameters.  Indeed, this is why the HB is ``horizontal''.  In some
sense it is then easier for a 2nd parameter candidate to shift
the HB morphology blueward.  In metal-rich GCs, however, the situation
is quite different.  Due to their high envelope opacity metal-rich HB
stars are normally confined to a red clump.  To move such stars
blueward requires a larger change in the stellar structure.  Thus
any 2nd parameter candidate capable of producing hot HB stars
in a metal-rich GC might also have other observational consequences
such as upward sloping HBs as in NGC 6388 and NGC 6441.
                
We have undertaken extensive stellar evolution calculations
in order to explore the cause of the 2nd parameter effect
in NGC 6388 and NGC 6441.  Some preliminary results from
SC98 are briefly reviewed in the following two sections.  We
first demonstrate that canonical HB models cannot explain the
upward sloping HBs in these GCs (section 2) and then present
three noncanonical scenarios based on 1) a high cluster helium
abundance, 2) internal rotation, and 3) helium mixing along
the RGB (section 3).  Finally in section 4 we discuss a new scenario
in which dust formation near the
tip of the RGB might lead to metal depletion in the envelopes
of the blue HB stars.

\section{Failure of Canonical HB Models}

We first consider whether canonical models can explain the upward
sloping HBs in NGC 6388 and NGC 6441.  It is well-known
that the effective temperature of an HB star increases as its
mass decreases.  Thus one could, in principle, produce hot
HB stars by increasing the cluster age and hence lowering the
main-sequence turnoff mass or by enhancing the amount of mass loss
along the RGB.  Rich et al. (1997), in fact, considered both of these
possibilities but found neither of them to be satisfactory.  More
specifically, the required increase in the cluster age would
be quite large, at least ${\approx} 4$ Gyr, which would place NGC 6388 and
NGC 6441 among the oldest, if not the oldest, GCs in the Galaxy.  Moreover,
the frequency of stellar interactions within the cores of these
clusters seems too low to produce the additional RGB mass
loss.  The lack of a radial gradient in the 
blue HB/red HB number ratio (Rich et al. 1997) also argues against
a dynamical explanation, although Layden et al. (1999) have suggested
that such a gradient may be present in NGC 6441.
                
In order to explore the canonical scenario in more detail, we have computed
grids of HB sequences for helium abundances $Y$ of 0.23 and 0.28 and
heavy-element abundances $Z$ of 0.002, 0.006 and 0.01716 (= $Z_{\sun}$).  The
tracks for one of these grids are plotted in
Figure 1.  Using such tracks, SC98 then computed HB simulations
\begin{figure*}[t]
\hspace{-0.49in}{\epsfbox{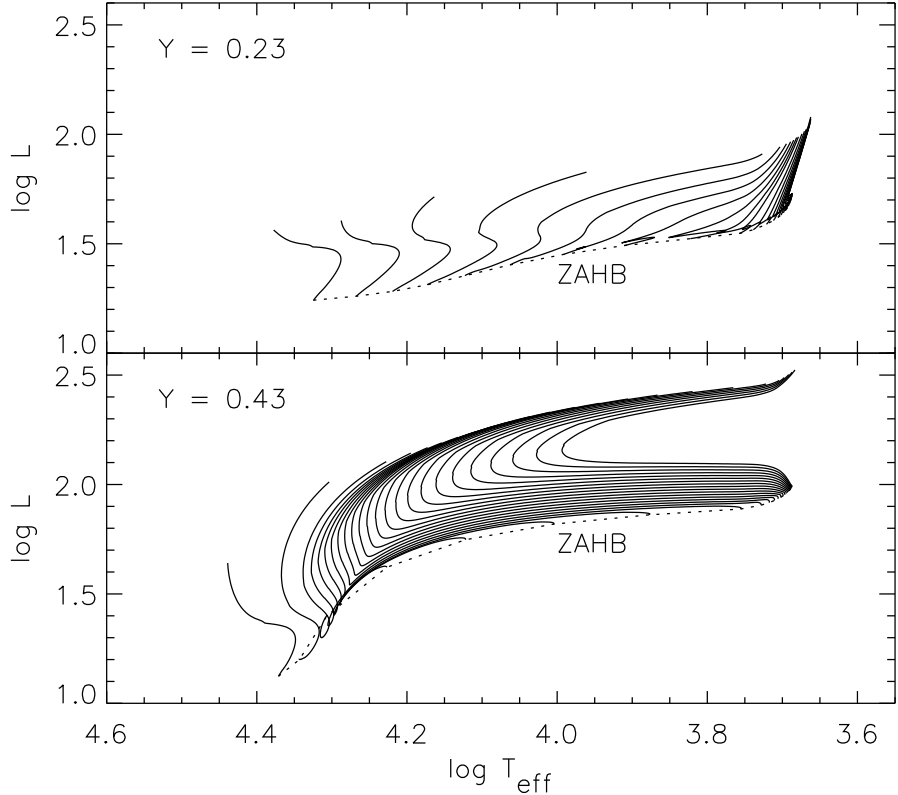}}
\caption{Canonical HB tracks for $Y$ = 0.23 (upper panel)
and high-$Y$ HB tracks for $Y$ = 0.43
(lower panel).  In both panels $Z$ = 0.006.  The zero-age HB (ZAHB)
is indicated by the dotted line.}
\label{figure 1}
\end{figure*}
in the (${M_V}$, {\bv}) plane, assuming a Gaussian distribution in
the HB mass (see Figure 1 of SC98).  In all of these simulations
the HB morphology was found to be quite flat without any indication
of an upward slope as observed in NGC 6388 and
NGC 6441.  Increasing
the cluster age or the RGB mass loss simply moved the models
blueward in {\bv} without increasing their luminosity.  We
conclude therefore that canonical HB models cannot account for the HB
morphology in NGC 6388 and NGC 6441.  In particular, the two
most prominent 2nd parameter candidates - age and RGB
mass loss - do not work.

\section{Noncanonical HB Scenarios}

The previous discussion indicates that the 2nd parameter
is increasing the luminosity of the blue HB stars in NGC 6388
and NGC 6441.  Theoretical models show that the HB luminosity
for a fixed $Z$ depends on two parameters: the helium abundance $Y$
and the core mass $M_{\rm c}$.  This fact prompted SC98
to suggest 3 noncanonical scenarios involving increases in
either $Y$ or $M_{\rm c}$ which might potentially produce
upward sloping HBs.  By ``noncanonical'' we mean scenarios which
include effects such as rotation or mixing that are normally
omitted in low mass stellar models or which assume nonstandard
values for the stellar parameters.  We will now briefly review
each of these scenarios and will then discuss some new observational
data which have become available since the SC98 study.
               
The first (``high-$Y$'') scenario assumes that the stars in NGC 6388
and NGC 6441 formed with a high initial helium abundance due to
a peculiar chemical enrichment history in these clusters.  From
theoretical models we know that high helium abundances
can lead to very long blue loops in the
HB tracks and to a large deviation of these tracks
from the zero-age HB (ZAHB).  To determine
if such track characteristics might produce
upward sloping HBs, we extended our canonical calculations
to $Y$ = 0.43 at increments
of 0.05.  One grid of high-$Y$ tracks is given in the
lower panel of Figure 1.  Comparing these high-$Y$ tracks with
the canonical tracks in Figure 1 shows the dramatic effect
that a high helium abundance can have on the track
morphology.  In the canonical case the blue loops are
quite short, with most of the evolution being spent near
the ZAHB.  At high helium abundances, however, the blue loops
extend to high effective temperatures, with the
blue end of the loops, where the evolution is particularly
slow, being much brighter than the ZAHB.  The HB simulations
given in Figures 2a,b of SC98 show that such high-$Y$ tracks
can indeed produce upward sloping HBs, but only if $Y$ is very large
(${\rm >}$ ${\approx} 0.4$).  Moreover, this scenario predicts a
very bright HB luminosity as well as a large value for
the number ratio $R$ of HB stars to RGB stars brighter than
the HB.
               
\begin{figure*}[t]
\hspace{-0.49in}{\epsfbox{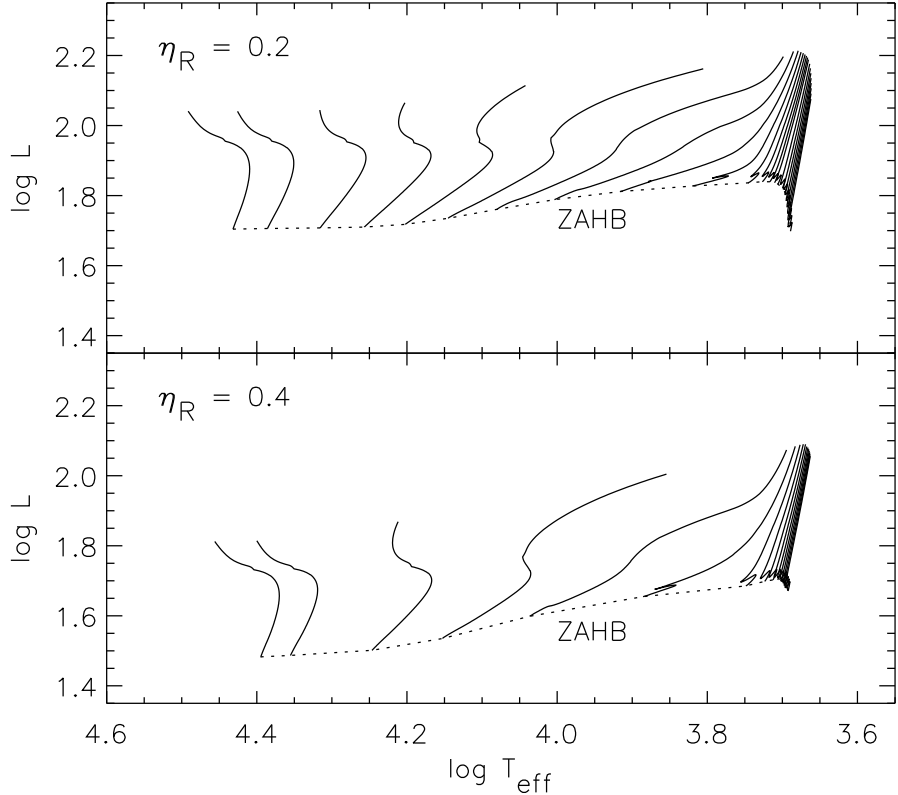}}
\caption{Rotation HB tracks
for a Reimers mass loss parameter $\eta_{\rm R}$ = 0.2
(upper panel) and 0.4 (lower panel).  In
both panels $Z$ = 0.006.  The ZAHB is indicated
by the dotted line.  The reddest track in each panel is a
canonical track without rotation.  The tracks shift
blueward as rotation increases the core mass
$M_{\rm c}$.}
\label{figure 2}
\end{figure*}
The second (``rotation'') scenario is based on the fact that
rotation within an RGB star can delay the helium
flash, thereby leading to a larger core mass and to a
brigher RGB tip luminosity and hence greater mass loss.  This
increase in $M_{\rm c}$ together with the corresponding decrease
in the mass $M$ then leads to a bluer and brighter HB morphology.  Thus
rotation might also be able to produce upward sloping
HBs, where the slow rotators are located in the
red clump while the fast rotators are shifted blueward
as the rotation rate increases.  In order
\begin{figure*}[t]
\hspace{-0.49in}{\epsfbox{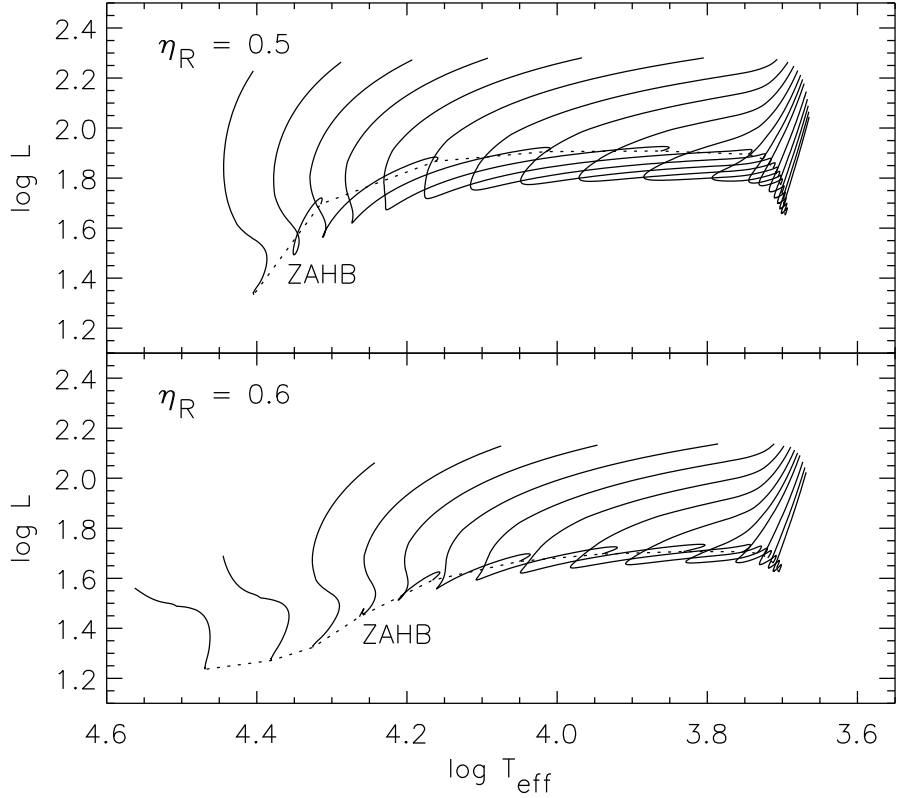}}
\caption{Helium-mixing HB tracks
for a Reimers mass loss parameter $\eta_{\rm R}$ = 0.5
(upper panel) and 0.6 (lower panel).  In
both panels $Z$ = 0.006.  The ZAHB is indicated
by the dotted line.  The reddest track in each panel
is a canonical track without helium mixing.  The
tracks shift blueward as the helium mixing increases.}
\label{figure 3}
\end{figure*}
to test this possibility, we need to know
how much extra mass is lost at the tip of the RGB when
$M_{\rm c}$ exceeds its canonical value.  To answer this question,
we evolved sequences up the RGB with the helium burning turned
off, thereby permitting $M_{\rm c}$ to become arbitrarily
large.  In this way we were able to determine how
$M$ decreases with increasing $M_{\rm c}$ for various
values of the Reimers mass loss parameter $\eta_{\rm R}$.  Using
these results to specify the variation of $M$ with $M_{\rm c}$
along the HB, we then computed grids of ``rotation'' tracks,
two of which are presented in Figure 2.  Perhaps not surprisingly,
the HB simulations obtained from these tracks
(see Figures 2c,d of SC98) also show upward sloping
HBs.  Thus rotation may offer an alternative explanation
for the HB morphology in NGC 6388 and NGC 6441.  Compared
to the high-$Y$ scenario, the rotation scenario predicts
smaller increases in both the HB luminosity and the $R$ ratio.
       
The third (``helium-mixing'') scenario is motivated by the large
star-to-star abundance variations in C, N, O, Na and Al which are found
among the red-giant stars within individual GCs and which are often
attributed in part to the mixing of nuclearly processed material
from the vicinity of the hydrogen shell out to the stellar
surface (Kraft 1994).  This mixing is believed to be driven by rotation.  The
observed enhancements in Al
are particularly important because they indicate that
the mixing is able to penetrate into the hydrogen shell (Cavallo et al.
1998), thereby dredging up fresh helium together with the Al.  Such
helium mixing would lead to an enhancement in the envelope
helium abundance during the RGB phase as well as to
additional mass loss due to a brighter RGB tip luminosity.  Thus
a helium-mixed star would arrive on the HB with both a higher
envelope helium abundance and a lower mass and therefore would be
both bluer and brighter than its canonical counterpart - just what
is needed to produce an upward sloping HB.  To test this possibility,
we have evolved many sequences up the RGB in which helium mixing
was included following the prescription of Sweigart (1997a,b).  Each
of these sequences was then
evolved through the helium flash and
HB phases.  Two HB grids computed in this manner
are given in Figure 3.  The
HB simulations obtained from these tracks show an upward slope
very similar to that seen in NGC 6388 and NGC 6441 (see Figures 2e,f
of SC98).  The predicted increases in both the HB luminosity
and the $R$ ratio are again smaller than in the high-$Y$ scenario.
      
All three of the above scenarios predict brighter blue HB stars
than canonical models.  This prediction has several interesting
consequences which we now examine in light of some recent
observational results (see also SC98).
       
First, from the pulsation equation we know that a brighter RR Lyrae
luminosity should lead to a longer pulsation period.  Recent
studies by Layden et al. (1999) and Pritzl et al. (1999)
have, in fact, found that the RR Lyrae variables in NGC 6388
and NGC 6441 are shifted towards longer periods in the period-amplitude
diagram compared to the metal-rich field RR Lyrae variables, just
as predicted by the noncanonical scenarios.  Moreover, the size of the
shift seems to be approximately consistent with some of the SC98
simulations, although a detailed analysis remains to be carried out.
        
Second, a brighter HB luminosity implies a larger value for the $R$
ratio.  Layden et al. (1999) have estimated an $R$ value of ${\approx} 1.6$
for NGC 6441.  While within the range expected from the
rotation and helium-mixing scenarios, this value is much smaller
than expected from the high-$Y$ scenario.
        
Third, one would expect the blue HB stars to have lower surface
gravities.  Moehler et al. (1999a) have recently obtained log $g$
values for 7 blue HB stars in NGC 6388 and NGC 6441.  Quite surprisingly,
these gravities are even larger on average than the
canonical gravities.  This is a baffling result for it says
that the noncanonical models which produce upward sloping HBs
cannot explain the gravities of the blue HB stars.  This
point is discussed further in the next section.
               
\section{Metal-Depletion Scenario}
                
When confronted with observations that cannot be explained by
canonical models, it is often helpful to examine the assumptions
that go into the calculations to see if they are fully justified.  One
assumption underlying both the canonical and noncanonical scenarios
discussed thus far is that the envelope heavy-element abundance
remains constant as a star evolves.  Suppose, however, that elements
with high condensation temperatures (e.g., Ca, Fe) were to form dust
grains which are then selectively removed from the stellar atmosphere
by radiation pressure near the tip of the RGB, thereby leaving
behind metal-depleted gas.  Suppose also that some of this gas
is then captured by the convective envelope of a red-giant
star, thus reducing the overall envelope metal abundance.  How
would such metal depletion affect the subsequent HB evolution?  If
the metals are depleted throughout the envelope, one would
expect an HB star to be both bluer and brighter.  Thus metal
depletion via dust formation near the tip of the RGB might
offer another way to produce upward sloping HBs.
               
Is there observational evidence for dust formation and metal
depletion in GC stars?  Moehler et al. (1998) have analyzed HST-GHRS
spectra of the two post-asymptotic-giant-branch (AGB) stars
Barnard 29 in M 13 and ROA 5701 in ${\omega}$ Cen.  Both stars
showed substantial Fe depletion
which Moehler et al. (1998) attributed to gas-dust separation
near the end of the AGB phase.  The CNO elements, however,
were not depleted.
               
A similar abundance pattern is also seen in field post-AGB stars.  In
particular, the Fe-group elements, especially Ca and Fe, often show
moderate to extreme deficiencies, with [Fe/H] reaching values below -4 in
some stars.  Bond (1991) has proposed that these ultra-low Fe abundances
arise from dust grain formation, a possibility that has been further
explored by Mathis \& Lamers (1992) (see also Napiwotzki et al. 1994).  Besides
CNO, a few other elements, most notably S, seem to have nearly
normal abundances, as would be expected under this metal-depletion
scenario.  Further support is provided by the presence
of dust around most post-AGB stars.
                  
Venn \& Lambert (1990) have emphasized the striking similarity between
the abundance pattern found in post-AGB stars and the patterns seen
in the interstellar gas and in a class of metal-deficient,
Population I A-type dwarfs known as ${\lambda}$ Bo\"{o}tis stars.  From an
analysis of 3 ${\lambda}$ Bo\"{o}tis stars, Venn \& Lambert (1990) concluded
that the strong depletions of Ca and Fe found in these stars together
with their approximately solar C, N, O and S abundances are
most likely due to dust
formation followed by the accretion of metal-depleted gas.  Significantly
Corbally \& Gray (1996)
have reported that 10 out of 67 of their field HB candidates also show
${\lambda}$ Bo\"{o}tis-like spectra including the prototype field HB
star HD 2857.  Again dust formation along the RGB was suggested as the most
likely explanation (Gray et al. 1996).  In addition, Adelman \& Philip (1994)
have noted that about one-half of their sample of field HB
stars had [Ca/Fe] values below those expected from metal-deficient
red giants.
       
There is also some evidence for Ca depletion in GC HB stars.  The
[Ca/H] distribution of the blue HB stars in ${\omega}$ Cen derived
by D'Cruz et al. (1997) contains a substantial population
of very Ca-poor stars that are not seen on the RGB.  Likewise,
the [Fe/H] distribution for the RR Lyrae variables in ${\omega}$ Cen
given by the ${\Delta} S$ method is skewed towards lower abundances
compared to the red giants.  We suggest that these HB
Ca depletions may be due to dust formation in the red-giant
progenitors.  Such an effect would complicate the use
of the ${\omega}$ Cen variables for determining the RR Lyrae
period shift even if the metal depletion was confined to the
surface layers and therefore had no effect on the luminosity.  In
general, a Ca depletion on the HB might lead to an offset between
the ${\Delta} S$ - [Fe/H] calibrations defined by the field and
cluster variables when giant branch [Fe/H] abundances are used.  Such
an offset has been reported by Carretta \& Gratton (1997).
                  
Dust formation and metal depletion
in bright GC red giants may be driven by stellar pulsations which
can lead to extended atmospheres and to large excursions in color
and hence effective temperature (see Fig. 6 of Montegriffo
et al. 1995).  Indeed, the connection between pulsation and mass
loss has often been noted (e.g., Frogel \& Elias 1988).  Several
types of bright red variables are present in GCs.  Metal-rich GCs
with [Fe/H] $>$ -1 contain long period variables (LPVs) having
luminosities brighter than the tip of the
RGB. These LPVs, which are presumably AGB stars, show substantial
infrared emission, indicating the presence of a circumstellar
dust shell (Frogel \& Elias 1988).  Intermediate-metallicity and
metal-rich GCs also contain ``peculiar'' variables
which are located just below the tip of the RGB
and which also show excess infrared emission.  Since the
evolutionary timescale near the RGB tip is ${\approx} 10^6$ yr,
one would expect only a few such variables per GC.  We suggest
that these peculiar variables may be the best candidates
for metal depletion.  NGC 6441 is, in fact, known to contain
a considerable population of red variables including Mira,
semi-regular and irregular variables (Layden et al. 1999).
               
Could metal depletion help to explain the high gravities of the
blue HB stars in NGC 6388 and NGC 6441 obtained by Moehler et al.
(1999a)?  Caloi (1999) was the first
to point out that an enhancement of the atmospheric metal abundances
in hot HB stars ($T_{\rm eff} > {\approx} 10,000 - 12,000$ K) by
radiative levitation might lead to an underestimate of the gravities
(see also Leone \& Manfr\`{e} 1997; Grundahl et al. 1999).  Subsequent
studies have confirmed both the existence of radiative levitation
in hot HB stars (Behr et al. 1999) and its effect on the derived
gravities (Moehler et al. 1999b).  In essence, if one analyzes
hot HB stars using the cluster metal abundance instead of
their actual (and larger)
atmospheric abundances, the gravities derived from the Balmer lines
will be too low.  Conversely, if the atmospheric abundances are lower
than the cluster abundance, then the gravities should
be too large.  Since all but 1 of the blue HB stars in the Moehler
et al. study had temperatures too cool for radiative
levitation, the large gravities might be partially understood if
the atmospheres of these stars were depleted of metals.
                
Metal depletion in bright red giants may also be relevant
to the current controversy between the ``long'' and ``short''
Population II distance scales.  Suppose that the surface
abundances of Ca and Fe are depleted in a GC red giant just
before it evolves from the tip of the RGB to the HB.  In
plotting such a star in the $M_V{\rm (HB)}$
versus [Fe/H] diagram, one would normally use the cluster [Fe/H]
abundance.  If, however, the same star were in the field,
one would then use its actual (and lower) surface abundance.  Since
$M_V{\rm (HB)}$ increases with increasing [Fe/H],
one would conclude that this field HB star is fainter than
cluster HB stars of the same metallicity
             
The above metal-depletion scenario is, of course,
highly speculative.  We are not suggesting that all
GC red giants experience metal depletion (see, e.g., Lambert
et al. 1992) but rather that metal depletion may under
some circumstances act as a 2nd parameter, possibly associated
with pulsation.  Effects such as
rotation or helium mixing which brighten the RGB tip luminosity
may increase the likelihood of pulsation and hence
metal depletion.
            
We can, however, offer two observational
tests.  First, the abundance pattern of the blue HB stars in
NGC 6388 and NGC 6441 should show normal CNO abundances but
depleted Ca and Fe.  A key diagnostic would be S, which should
have the normal cluster abundance.  Second,
${\Delta} S$ measurements of the RR Lyrae variables
should indicate a lower than cluster [Fe/H] abundance.  Finally
theoretical calculations are needed to see if metal depletion
with the expected abundance pattern can, in fact, lead to upward
sloping HBs.

\acknowledgments

The author gratefully acknowledges the many contributions made to
this research by his collaborators M. Catelan, A. Layden, S. Moehler,
B. Pritzl, R. Rich, \& H. Smith.

\end{document}